\documentclass[10pt,a4paper]{article}

\usepackage{amsmath}
\usepackage{amsfonts}
\usepackage{latexsym}
\usepackage{graphicx}

\begin{document}

\title{\Large Research at UNIS - The University Centre in Svalbard. A bibliometric study}
\author{Johannes Stegmann\footnote{Member of the Ernst-Reuter-Gesellschaft der Freunde, F\"orderer und Ehemaligen der Freien Universit\"at Berlin e.V., Berlin, Germany, johannes.stegmann@fu-berlin.de} }
\date{} 
\maketitle

\begin{abstract} The scientific output 1994-2014 of the University Centre in Svalbard (UNIS) was bibliometrically analysed. It was found that the majority of the papers have been published as international cooperations and rank above world average. Analysis of the papers' content reveals that UNIS works and publishes in a wide variety of scientific topics.  
\end{abstract}

\section{Introduction} \indent
\indent The sensitivity of the Arctic to climate changes and the heavy impact of such transformations on other world regions (Post et al., 2009) as well as its presumed richness in oil, gas and other mineral deposits has moved the Arctic into the focus of intensive scientific, economic, political and public attention (Humrich, 2013).\\
\indent The University Centre in Svalbard (UNIS) was established in 1993 as  "Arctic extension" of Norway's universities 
(UNIS, 2009 a). UNIS is to "represent and secure Norwegian polar interests" (UNIS, 2009 a). UNIS's mission is also to offer an international research platform for all kinds of basic Arctic research (UNIS, 2009 b). \\
\indent It seems to be of interest to analyse UNIS' scientific acitivities from a bibliometric point of view. This communication tries to answer the following questions: (i) What is produced by UNIS in terms of scientic papers? (ii) To what extent is UNIS' propagated internationality realised in terms of international coauthorships? (iii) What is the standing of UNIS' publications in terms of appropriate international standards? (iv) What is the content of UNIS authored papers in terms of subfields and subject topics?

\section{Methods} \indent
\indent Papers published since 1994 by UNIS were retrieved and downloaded from the Web of Science (WoS) on January 19, 2014, using an appropriate address search profile.  \\
\indent For the  analysis of UNIS' paper output and its distribution to different document types all retrieved records were used. For the analysis of UNIS' research those papers not being research articles (i.e. not of document type "ARTICLE") were excluded. \\
\indent The citation performance of UNIS' papers was measured applying the Percentile Rank Index (PRI) developed by Pudovking and Garfield (Pudovkin and Garfield, 2009). I call this version of a percentile rank index PG-PRI but use in this paper "PG-PRI" and "PRI" synonymously because no other PRI methods are involved here. \\
\indent Prior to PG-PRI calculation of a paper in question the citation rank of this paper among its "paper peers", i.e. all papers published in the same source journal in the same year must be determined. Because most papers need some time to gather cites it makes no sense to include too recent papers in a PRI analysis. In this study, only research papers (document type "ARTICLE") of UNIS published before 2013 (i.e. published in the years 1994-2012) were included (723 papers). For each of these 723 research articles its publication year and publishing journal was determined, and all papers (document type "ARTICLE" only) of the corresponding journal-year pair were retrieved and downloaded. In summary, the papers of 514 journal-year pairs were retrieved and downloaded between the 6\textsuperscript{th} and 10\textsuperscript{th} February 2014. Then, the papers of each journal-year set were ranked top-down according to citations received. In case of ties (several papers having the same citation frequency), each of the tied values was assigned the average of the ranks for the tied set (Pudovkin and Garfield, 2009, Pudovkin et al., 2012). The position of each of the UNIS papers in the corresponding paper set was determined. PG-PRI values were calculated according to the formula 
\begin{displaymath}
PRI = \frac{N-R+1}{N}*100 
\end{displaymath}
where N is the number of papers in the year set of the journal and R is the citation rank of the paper (Pudovkin and Garfield, 2009). R=1 is the top rank (most cited paper) with PRI=100 (Pudovkin and Garfield, 2009).\\
For determination of the global (expected) average PRI the Svalbard papers were ordered according to the number of papers published in the corresponding journal-year set. The average PRI was calculated according to the formula
\begin{displaymath}
PRI_{globav} = 50 + \frac{50}{N}.
\end{displaymath}
where N is the number of papers published in the journal-year pair at the median position of the ordered set (Pudovkin et al., 2012). In the present study, the median N was found to be 150; therefore, 
\begin{displaymath}
PRI_{globav} = 50.33
\end{displaymath} \\ 
\indent For cluster analysis of keywords the co-word analysis technique described by Callon et al. (1991) was applied. A detailed description of the algorithm can be found in Stegmann and Grohmann (2003).\\
\indent Extraction of record field contents, clustering, data analysis and visualisation were done using homemade programs and scripts for perl (version 5.14.2) and the software package R version 2.14.1 (R Core Team, 2013). All operations were performed on a
commercial PC run under Ubuntu version 12.04 LTS.
\newpage

\section{Results and Discussion} 
\subsection{Output (papers)} \indent
\indent 
Since 1994 UNIS published 875 papers, more than 85\% of them being research papers (Figure 1). In UNIS' starting years only few papers were published but the annual publication numbers gradually increased up to 94 in 2012. In 2013 only 73 UNIS papers were retrieved but probably not yet all papers with 2013 as publication year had been recorded to the WoS database at the time of retrieval (Januar 2014). The number of UNIS papers retrieved from the WoS database are in good agreement with the corresponding numbers derived from UNIS' annual reports 2009 to 2012 (UNIS 2009 b, 2010, 2011, 2012). For the subsequent analysis of Svalbard's research papers of document type "ARTICLE" only were included. These amount to 748 papers for the whole time span (1994 - January 2014). These papers have 2331 distinct authors; the most prolific author is (co)author of 66 papers. The average number of authors per paper is 3.1; predominant is the class with 4 authors per paper. Only 4.1\% (31 papers) of the research papers are single-authored (not shown). The paper with the highest number of authors (376) is the yearly published  {\em State of the Climate} report, a special supplement to the {\em Bulletin of the American Meteorological Society} (Blunden and Arndt, 2012).\\
\indent 67\% of UNIS's research papers are international papers, jointly authored by at least one author of UNIS (i.e. from Norway) and at least one author from another country. In total, 56 different countries (including Norway) are involved in UNIS' research papers. Table 1 shows the top 15 cooperating countries. Among them are the other (besides Norway) circumpolar countries: Canada, Denmark (due to its autonomous region Greenland), Iceland, Russia, USA. \\
\indent UNIS has published its papers in more than 200 journals; the top ten are displayed in Table 2 (see also Table 3 and next section).
 
\subsection{Benchmarking (PG-PRI)} \indent 
\indent 
For the analysis of the international standing of UNIS' research the percentile rank indexing method of Pudovkin and Garfield (2009) was applied (PG-PRI, see Methods). Figure 2 displays the PG-PRI value of each of the 723 research articles of UNIS. Table 4 lists some PRI ranges. The average PRI value of UNIS' 723 research articles published 1994-2012 is 53.9, well above the expected (global) mean of 50.33 (see Methods). In addition, more than one half (392 = 54.2\%) of the UNIS papers have PG-PRI values above the global mean (Figure 2, Table 4). The PG-PRI has the inherent capability for international comparison of an author's/institute's papers because it compares the citation performance of the research papers in question with their "direct peers", i.e. papers of the same type published in the same journals in the same time span ((Pudovkin and Garfield, 2009, Pudovkin et al., 2012). From the data in Figure 2 and Table 4 it is concluded that UNIS' research perform well above the average of comparable world research. This conclusion is supported by the high impact factor ranks of the top ten journals with UNIS papers within their JCR categories (see Table 3).

\subsection{Content (categories, keywords)} \indent 
\indent 
Rough indicators of the scientific (sub)fields to which papers contribute are the WoS categories to which journals are assigned. UNIS contributes to 52 WoS categories. The top 15 categories to which UNIS research papers (i.e. the publishing journals) have been assigned are shown in Table 5. Earth, marine, environmental sciences play a role, but also space and evolutionary sciences are important.  \\
\indent Deeper insights into UNIS' research areas may be achieved by an analysis of the keywords assigned to the articles. The keywords were extracted from the record fields DE (author keywords) and ID (keyword plus). 3999 distinct keywords were extracted and - in a first step - assigned to WoS categories of the respective articles. Identical keywords (occuring in both fields, DE and ID) were counted only once. \\
\indent Table 6 lists for the 10 top categories (see Table 5) frequent keywords (omitting not so informative descriptors like Svalbard, Spitsbergen, Sea, etc.). \\
\indent Another possibility to get an overview of the content of a set of papers is cluster analysis of the relevant keywords. Here, the co-word analysis of Callon et al. (1991) was applied (see Methods). The perl scripts developed by Stegmann and Grohmann (2003) were used to perform a cluster analysis of the keywords assigned to UNIS' research articles. The keywords are clustered according to their mutual similarity based on their co-occurrence strength, determined by calculation of the {\em cosine}. Several thresholds were applied: (i) minimal frequency = 4 (occuring of a keyword in at least 4 records, this threshold reduced the number of keywords to 395); (ii) minimal cosine similarity = 0.2; (iii) mininal (maximal) cluster size = 3 (10), i.e. only clusters containing at least 3 keywords were included. The upper threshold of cluster size was applied in order to avoid clusters with very many keywords which makes the cluster readability difficult. When a cluster had accumulated 10 terms a new cluster was started (see Stegmann and Grohmann, 2003). \\
\indent Under these restrictions 38 clusters with 338 keywords were found. For their visualisation the form of a density-centrality diagram (Callon et al., 1991) was chosen. Density means the strength of the links between cluster members, centrality means the strength of links between clusters. To label a cluster (besides its number) the keywords contained in it were ranked according to the product of link strengths (cosine values) and frequency; the keyword with the highest product was selected as cluster name. \\
\indent The diagram is displayed in Figure 3. Theoretically, this kind of diagram (also called "strategical diagram", see Callon et al., 1991) displays in its right half "central" topics. i.e. subjects with strong or many links to other topics (clusters). In the upper half the topics are "dense", i.e. the keywords constituting a topic are mutually strongly linked but do not have necessarily many or strong links outside the cluster. In the left and lower parts of the diagram subjects developing to more centrality and/or density are displayed. \\
\indent Tables 7, 8, 9, 10 lists all clusters with their keywords. Table 7 lists the keywords contained in the clusters positioned in the lower left quadrant of the diagram, Table 8 those of the upper left quadrant, and in Tables 9 and 10 the keywords of the lower and upper right quadrants are tabulated. Inspection of the clusters clearly shows that in most cases co-clustered keywords express associated topics. For a non-expert, however, it is difficult to make substantial statements concerning the cluster contents and the position of the research represented by the cluster keywords within the frame of UNIS's research. Of course, the help of field experts is needed for a detailed analysis of topics and their keywords. Variation (lowering) of the applied thresholds could result in a more detailed view of topics dealt with by UNIS' research papers. One should, however, not forget that the keywords in WoS records do not belong to a controlled vocabulary. It might happen that the same issue is described with (slightly) different keywords in different records, thus lowering the link strengths from this issue to other ones. Nevertheless, the data displayed in Figure 3 and Tables 7 to 10 show that UNIS has a diversified spectrum of Arctic research topics. 

\section{Conclusion} \indent
\indent The University Centre in Svalbard (UNIS) has been found to be a high-level research centre performing well above world average in wide variety of research subjects.

\section{Acknowledgements} \indent
\indent Part of this paper has been submitted for poster presentation at the 15\textsuperscript{th} COLLNET meeting 2014.\\ \indent Helpful PRI-related comments of Alexander Pudovkin are gratefully acknowledged.

\begin{figure}[hbtp]
  \centerline{
  \includegraphics{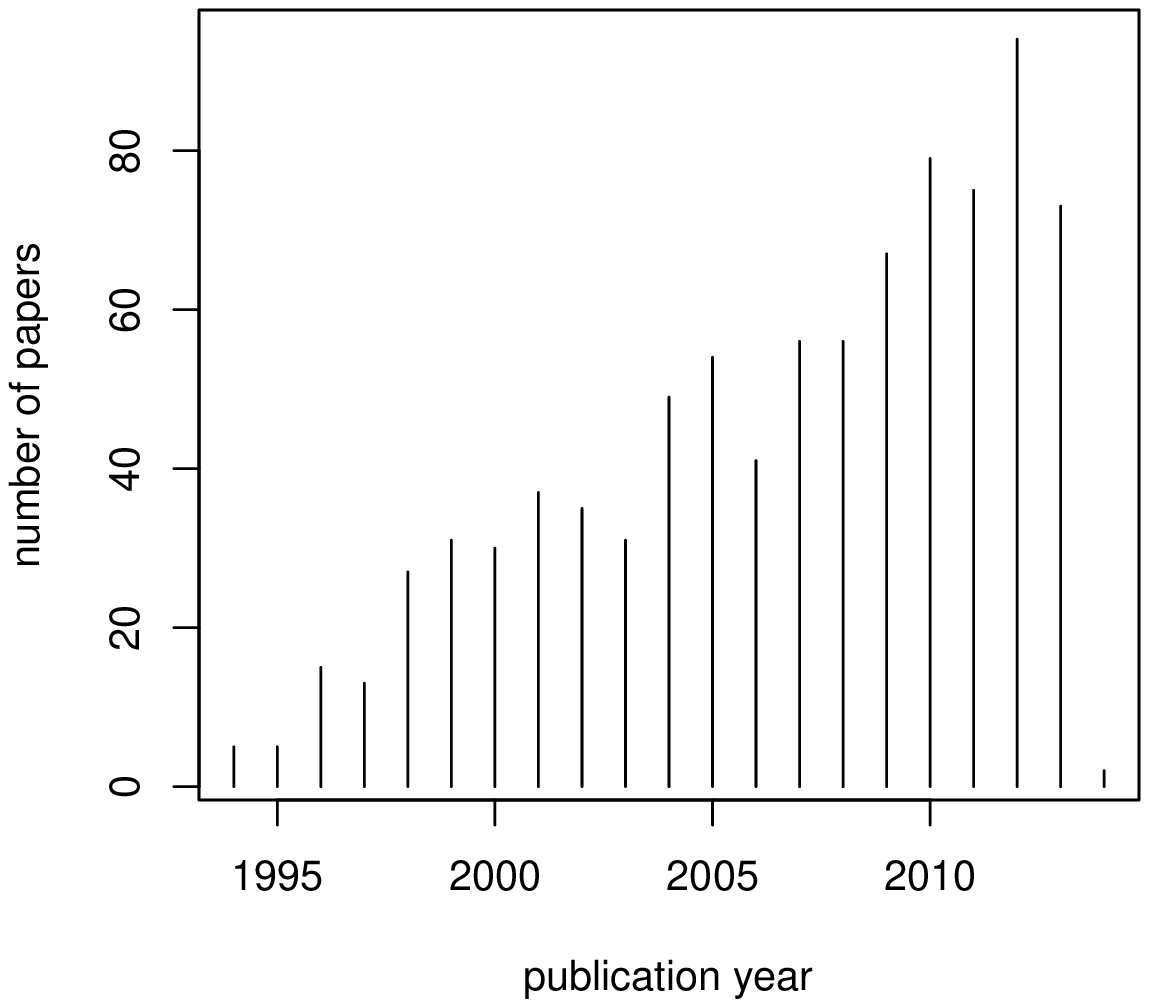}}
    \caption{Publications of UNIS 1994-2014\textsuperscript{*}.\newline
   \textsuperscript{*}Total no. of papers: 875. \newline
   (85.5\% articles, 5.9\% proceedings papers, 5.6\% reviews)
 }
  \label{fig:fig01}
\end{figure}

\begin{figure}[hbtp]
  \centerline{
  \includegraphics{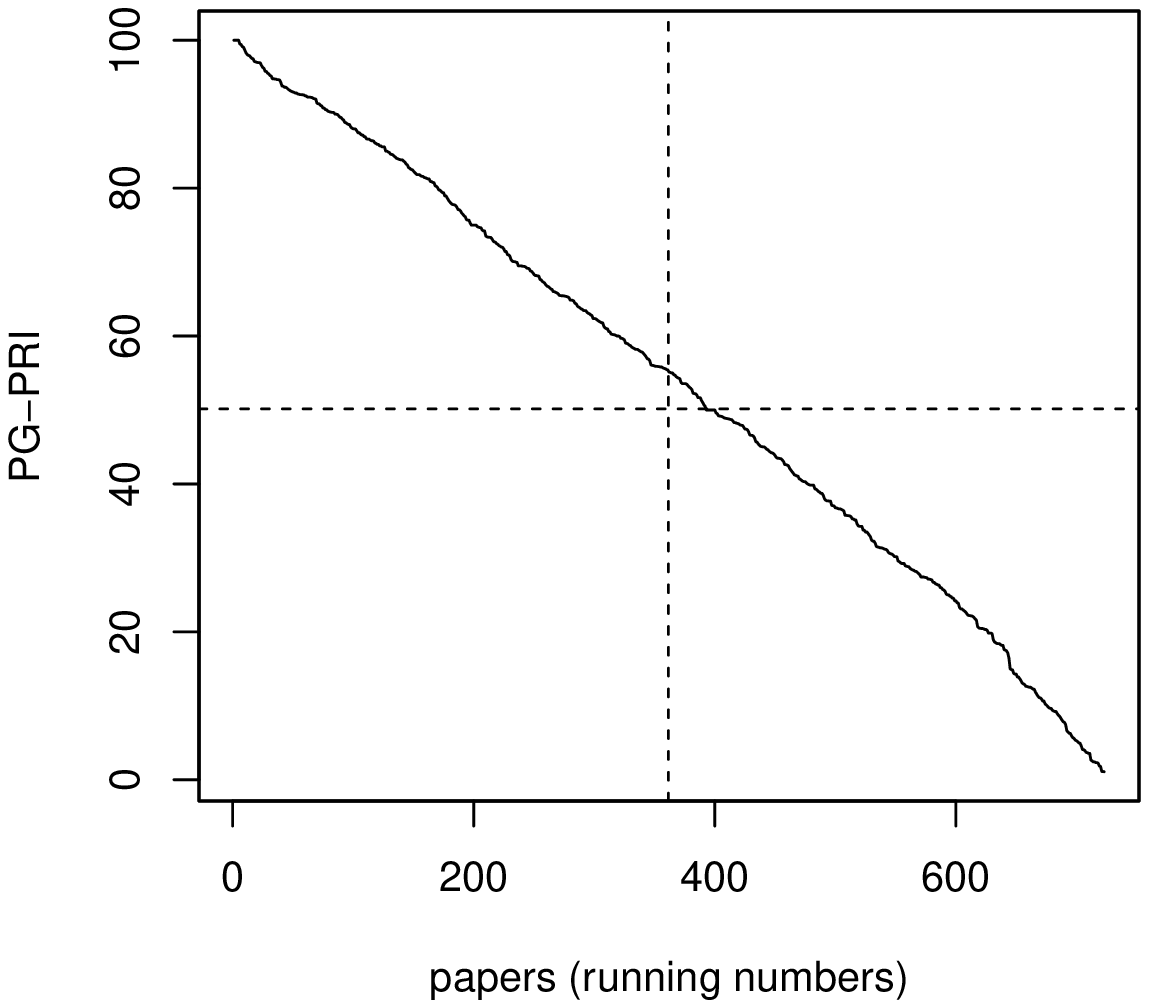}}
    \caption{Percentile Rank Indexes of UNIS' research papers 1994-2012\textsuperscript{*}.\newline
   \textsuperscript{*}Total no. of research papers: 723.\newline
     PG-PRI: Pudovkin-Garfield Percentile Rank Index (see Methods).\newline
     Vertical dashed line: median of papers.\newline
     Horizontal dashed line: expected global mean PRI (see Methods).
 }
  \label{fig:fig02}
\end{figure}

\begin{figure}[hbtp]
  \centerline{
  \includegraphics{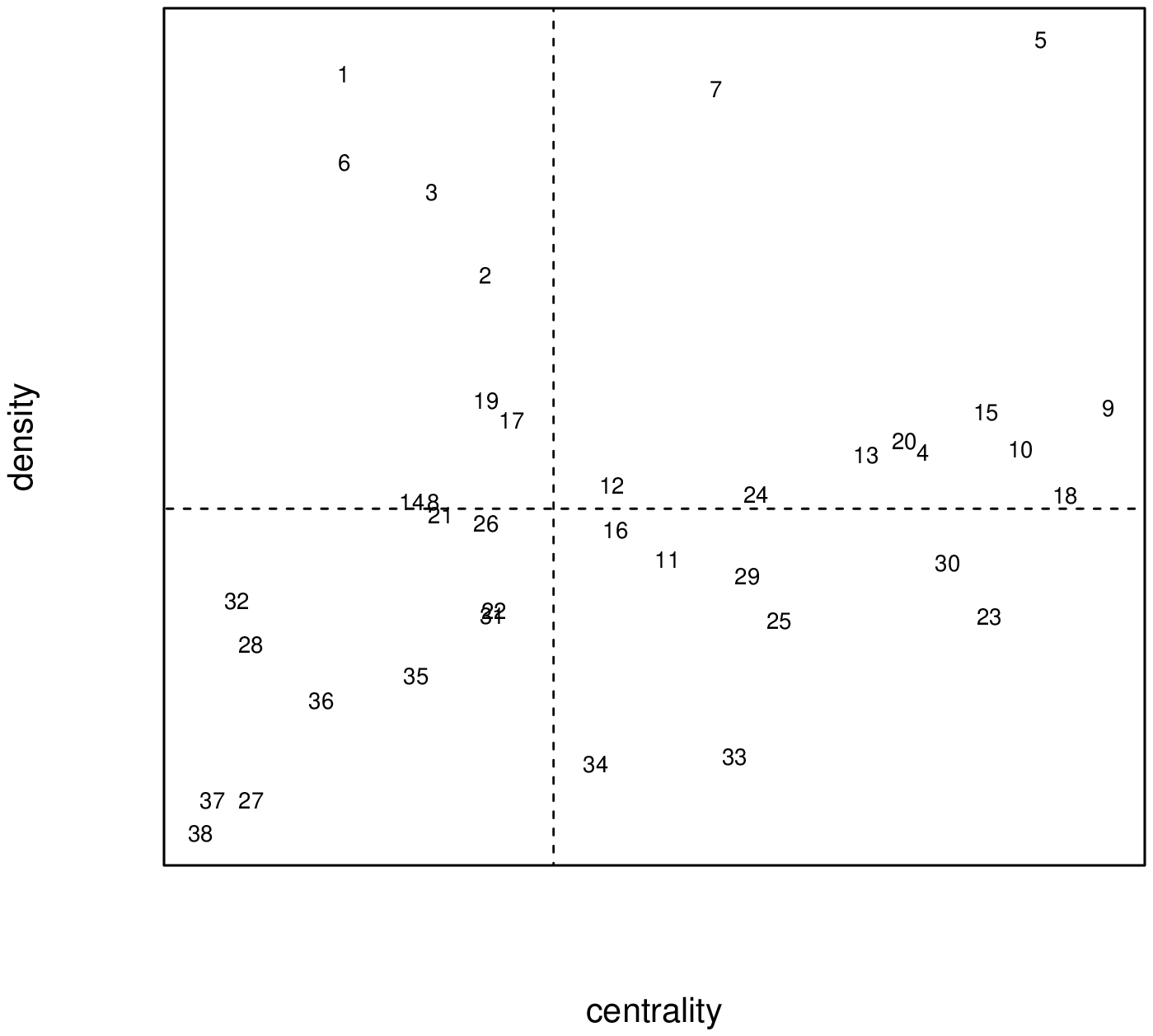}}
    \caption{UNIS' research papers 1994-2014: centrality-density diagram of keywords clusters.\newline
     Cluster numbers are centered at cluster positions.\newline
     Centrality increases from left to right, Density increases bottom-up.\newline
     Vertical dashed line: median centrality.\newline
     Horizontal dashed line: median density.
 }
  \label{fig:fig03}
\end{figure}

%
%
%
%
\begin{table}
\centering
\caption{UNIS' research papers 1994-2014: cooperating countries (top 15).\newline
               }
%
%
\begin{tabular}{lccc}
\noalign{\smallskip}
\hline\noalign{\smallskip}
Country & No. of papers & \% of total (748 papers) & \\
\noalign{\smallskip} 
\noalign{\smallskip}\hline\noalign{\smallskip}
UNITED KINGDOM   & 197  &  26  & \\
 \noalign{\smallskip} 
USA   & 146   & 20 &  \\
\noalign{\smallskip} 
DENMARK  & 74 & 10 &  \\
\noalign{\smallskip} 
SWEDEN  & 66 & 9 &  \\
\noalign{\smallskip} 
GERMANY & 65 & 9 &  \\
\noalign{\smallskip} 
RUSSIA  & 46 & 6 &  \\
\noalign{\smallskip} 
POLAND & 40 & 5 &  \\
\noalign{\smallskip} 
CANADA  & 39 & 5  &  \\
\noalign{\smallskip} 
JAPAN  & 30 & 4  &  \\
\noalign{\smallskip} 
FRANCE  & 29 & 4  &  \\
\noalign{\smallskip} 
FINLAND  & 24 & 3  &  \\
\noalign{\smallskip} 
ICELAND & 21 & 3  &  \\
\noalign{\smallskip} 
SWITZERLAND & 19 & 3  &  \\
\noalign{\smallskip} 
NETHERLANDS & 18 & 2  &  \\
\noalign{\smallskip} 
ITALY  & 17 & 2  &  \\
\noalign{\smallskip}\hline
\end{tabular}
\end{table}
\begin{table}
\centering
\caption{UNIS' research papers 1994-2014: publishing journals (top 10).\newline
Number and fraction of papers published. 
           }
%
%
\begin{tabular}{lccc}
\noalign{\smallskip}
\hline\noalign{\smallskip}
Journal & No. of papers & \% of total (748 papers) &   \\
\noalign{\smallskip} 
\noalign{\smallskip}\hline\noalign{\smallskip}
J GEOPHYS RES  & 54 & 7 &   \\
\noalign{\smallskip} 
POLAR BIOL   & 50  &  7  &  \\
\noalign{\smallskip} 
GEOPHYS RES LETT   & 41   & 6 &    \\
\noalign{\smallskip} 
COLD REG SCI TECHNOL  & 25 & 3 &    \\
\noalign{\smallskip} 
POLAR RES & 24 & 3  &  \\
\noalign{\smallskip} 
J GLACIOL  & 21 & 3  &  \\
\noalign{\smallskip} 
ANN GEOPHYS GERMANY & 19 & 3 &  \\
\noalign{\smallskip} 
QUATERNARY SCI REV  & 19 & 3 &  \\
\noalign{\smallskip} 
MAR ECOL PROG SER  & 13 & 2 &  \\
\noalign{\smallskip} 
BOREAS  & 12 & 2 &  \\
\noalign{\smallskip}\hline
\end{tabular}
\end{table}
\begin{table}
\centering
\caption{UNIS' research papers 1994-2014: publishing journals (top 10).\newline
JCR Category and IF rank (JCR: Journal Citation Reports, IF: Impact Factor). \newline
In parentheses: numbers of journals in the category.
           }
%
%
\begin{tabular}{lcll}
\noalign{\smallskip}
\hline\noalign{\smallskip}
Journal &  IF 2012 &  JCR Category  & IF rank   \\
\noalign{\smallskip} 
\noalign{\smallskip}\hline\noalign{\smallskip}
J GEOPHYS RES  & 3.17 & GEOSCIENCES, & 23 (172)  \\
           &      & MULTIDISCIP.  &     \\
\noalign{\smallskip} 
POLAR BIOL & 2.01 & BIODIVERSITY & 14 (40)  \\
           &      & CONSERVATION &            \\
\noalign{\smallskip} 
GEOPHYS RES LETT   & 3.98  & GEOSCIENCES, & 11 (172)   \\
           &      & MULTIDISCIP.  &       \\
\noalign{\smallskip} 
COLD REG SCI & &  & \\  
TECHNOL     & 1.29 &  ENGINEERING, & 32 (122)   \\
           &      & CIVIL  &      \\
\noalign{\smallskip} 
POLAR RES & 1.62 & OCEANOGRAPHY & 30 (60)   \\
\noalign{\smallskip} 
J GLACIOL  & 2.88 & GEOGRAPHY,  & 12 (45)   \\
           &      & PHYSICAL  &       \\
\noalign{\smallskip} 
ANN GEOPHYS & & & \\
GERMANY & 1.52 & ASTRONOMY \&  & 31 (56)   \\
           &      & ASTROPHYSICS  &       \\
\noalign{\smallskip} 
QUATERNARY SCI REV  & 4.08 & GEOGRAPHY, & 3 (45)   \\
           &      & PHYSICAL  &       \\
\noalign{\smallskip} 
MAR ECOL PROG SER   & 2.55 & MARINE \&  & 16 (100)  \\
           &      & FRESHWATER BIOL.  &      \\
\noalign{\smallskip} 
BOREAS & 2.46 & GEOGRAPHY, & 18 (45)  \\
           &      & PHYSICAL  &        \\
\noalign{\smallskip}\hline
\end{tabular}
\end{table}
\begin{table}
\centering
\caption{UNIS' research papers 1994-2012: PG-PRI ranges.\newline
       \newline
    }
%
%
\begin{tabular}{lcc}
\noalign{\smallskip}
\hline\noalign{\smallskip}
PRI range &  No. of papers &  \% of total (723 papers)    \\
\noalign{\smallskip} 
\noalign{\smallskip}\hline\noalign{\smallskip}
PRI = 100 & 5 & 0.7  \\
\noalign{\smallskip} 
PRI $\ge$ 99 & 10 & 1.4 \\
\noalign{\smallskip} 
PRI $\ge$ 90 & 86 & 11.9  \\
\noalign{\smallskip} 
PRI $\ge$ 75 & 202 & 27.9   \\
\noalign{\smallskip} 
PRI $\ge$ 50.33 & 392 & 54.2   \\
\noalign{\smallskip} 
\noalign{\smallskip}\hline
\end{tabular}
\end{table}
\begin{table}
\centering
\caption{UNIS' research papers 1994-2014: WoS Categories (top 15).\newline
               }
%
%
\begin{tabular}{lccc}
\noalign{\smallskip}
\hline\noalign{\smallskip}
Category & No. of papers & \% of total & \\
         &               & (748 papers) & \\
\noalign{\smallskip} 
\noalign{\smallskip}\hline\noalign{\smallskip}
GEOLOGY   & 289  &  38  & \\
 \noalign{\smallskip} 
ENVIRONMENTAL    &   &  &  \\
SCIENCES \& ECOLOGY & 212  & 28 & \\
\noalign{\smallskip} 
PHYSICAL GEOGRAPHY  & 103 & 14 &  \\
\noalign{\smallskip} 
OCEANOGRAPHY  & 97 & 13 &  \\
\noalign{\smallskip} 
ASTRONOMY \& ASTROPHYSICS & 82 & 11 &  \\
\noalign{\smallskip} 
METEOROLOGY  &  &  &  \\
\& ATMOSPHERIC SCIENCES & 75 & 10 & \\
\noalign{\smallskip} 
MARINE  &  &  &  \\
\& FRESHWATER BIOLOGY & 66 & 9 & \\
\noalign{\smallskip} 
BIODIVERSY \& CONSERVATION  & 65 & 9  &  \\
\noalign{\smallskip} 
ENGINEERING  & 43 & 6  &  \\
\noalign{\smallskip} 
GEOCHEMISTRY \& GEOPHYSICS  & 30 & 4  &  \\
\noalign{\smallskip} 
ZOOLOGY  & 28 & 4  &  \\
\noalign{\smallskip} 
PLANT SCIENCES & 22 & 3  &  \\
\noalign{\smallskip} 
EVOLUTIONARY BIOLOGY& 21 & 3  &  \\
\noalign{\smallskip} 
LIFE SCIENCES \& BIOMEDICINE &  &   &  \\
- OTHER TOPICS  & 15  & 2 & \\
\noalign{\smallskip} 
PALEONTOLGY  & 13 & 2  &  \\
\noalign{\smallskip}\hline
\end{tabular}
\end{table}
\begin{table}
\centering
\caption{UNIS' research papers 1994-2014: Category-specific keywords.\newline
    }
%
%
\begin{tabular}{lll}
\noalign{\smallskip}
\hline\noalign{\smallskip}
WoS Category & Keywords & \\
\noalign{\smallskip} 
\noalign{\smallskip}\hline\noalign{\smallskip}
GEOLOGY   & \footnotesize{Ionosphere; Interplanetary Magnetic Field; Evolution;} & \\
   & \footnotesize{Permafrost; Mass Balance; Climate Change; Ice Sheet}  & \\
\noalign{\smallskip} 
ENVIRONMENTAL   &    &  \\
SCIENCES \& & & \\
ECOLOGY & \footnotesize{Barents Sea; Climate Change; Population Dynamics;} & \\
        & \footnotesize{Zooplankton; Plant Phenology; Cervus Elaphus;}  & \\
        & \footnotesize{Life History; Growth; Red Deer; Svalbard Reindeer;} & \\
        & \footnotesize{Large Herbivores; Vegetation} & \\
\noalign{\smallskip} 
PHYSICAL & & \\
GEOGRAPHY  & \footnotesize{Permafrost; Mass Balance; Climate Change; Glacier;} & \\
        & \footnotesize{Evolution; Alaska; Ice Sheet; North Atlantic; Sediments;} & \\
        & \footnotesize{Greenland Ice Sheet; Arctic Ocean} &  \\
\noalign{\smallskip} 
OCEANOGRAPHY  & \footnotesize{Barents Sea; Arctic Ocean; Interannual Variability;} & \\
        & \footnotesize{Sea Ice; Fram Strait; Circulation; Marginal Ice Zone;} & \\
        & \footnotesize{Atlantic Water} &  \\
\noalign{\smallskip} 
ASTRONOMY \& &  & \\
ASTROPHYSICS & \footnotesize{Interplanetary Magnetic Field; Ionosphere;} & \\
 & \footnotesize{Plasma; Particle Precipitation; Magnetosphere;} & \\
        & \footnotesize{Solar Wind; F Region; Fux Transfer Events;} &  \\
        & \footnotesize{Convection; Magnetopause} &  \\
\noalign{\smallskip} 
METEOROLOGY \& &  &    \\
ATMOSPHERIC & & \\
SCIENCES & \footnotesize{Interplanetary Magnetic Field; Polar Ionosphere;} & \\
        & \footnotesize{Temperature; Magnetospheric Physics; Atmosphere;} & \\
        & \footnotesize{Particle Precipitation; Ground Based Observations;}  & \\
        & \footnotesize{Convection; Thermosphere; Magnetopause} & \\
\noalign{\smallskip} 
MARINE \&  &  &    \\
FRESHWATER & & \\
BIOLOGY & \footnotesize{Barents Sea; Marginal Ice Zone; Zooplankton; Calanus;} & \\
        & \footnotesize{Vertical Migration; Community Structure; Kongsfjorden;} & \\
        & \footnotesize{Oxidative Stress; Climate Change; Population Dynamics} & \\
\noalign{\smallskip} 
BIODIVERSY \& & & \\
CONSERVATION  & \footnotesize{Barents Sea; Climate Change; Abundance;} & \\
        & \footnotesize{Zooplankton; Sea Ice; Life History; Growth; Diversity;} & \\
        & \footnotesize{Diet; Fauna; Kongsfjorden; Crustacea; Macrobenthos}  &  \\
\noalign{\smallskip} 
ENGINEERING  & \footnotesize{Sea Ice; Arctic; Consolidation; Field Experiments;} & \\
        & \footnotesize{Model; Strength; Glacier; Tow Line Tension;} & \\
        & \footnotesize{Thermal Regime; Air; Ice Ridges; Ice Rubble} &  \\
\noalign{\smallskip} 
GEOCHEMISTRY \& & & \\
GEOPHYSICS  & \footnotesize{Iceberg Discharges; Atmosphere; Ionosphere;} & \\
        & \footnotesize{Airglow; Greenland Ice; Evolution; Diffusion;} & \\
        & \footnotesize{Heinrich Events; Thermosphere; Nordic Seas;} & \\
        & \footnotesize{New Zealand; Ocean Circulation; Barents Shelf} &  \\
\noalign{\smallskip}\hline
\end{tabular}
\end{table}

\begin{table}
\centering
\caption{UNIS' research papers 1994-2014: content (keywords) of clusters positioned in the lower left quadrant (below median density, below median centrality) of the centrality-density diagram (see Figure 3). \newline
Italic terms: cluster label terms (the cluster keyword with the highest product of link strengths and frequency, see text). }
%
\begin{tabular}{lll}
\noalign{\smallskip}
\hline\noalign{\smallskip}
no. & keywords & \\
\noalign{\smallskip} 
\noalign{\smallskip}\hline\noalign{\smallskip}
21 & \footnotesize{{\em radar}; mesopause; mesosphere; pmse; high latitudes; summer mesosphere;} & \\
  & \footnotesize{particles; backscatter; tromso; noctilucent clouds} & \\ 
\noalign{\smallskip}  
26 & \footnotesize{{\em sediments}; airglow; middle; atmosphere; air temperature; latitudes;} & \\
  & \footnotesize{hf radar; fluctuations; series; frequency} & \\ 
\noalign{\smallskip}  
32 & \footnotesize{{\em diel vertical migration}; calanus; calanus finmarchicus} & \\ 
\noalign{\smallskip}
22 & \footnotesize{{\em mass balance}; age calibration; push moraines; deglaciation; tibetan plateau;} & \\
  & \footnotesize{retreat; valley glacier; structural evolution; sheet; southern norway} & \\ 
\noalign{\smallskip}
31 & \footnotesize{{\em range shifts}; migration; colonization; conservation; western svalbard;} & \\
  & \footnotesize{genetic diversity; spring bloom; dispersal; last glacial maximum;} & \\
  & \footnotesize{ eastern svalbard} & \\ 
\noalign{\smallskip}
28 & \footnotesize{{\em ground penetrating radar}; tasman glacier; swiss alps; new zealand;} & \\
   & \footnotesize{british columbia} & \\ 
\noalign{\smallskip}
35 & \footnotesize{{\em egg production}; vertical migration; west greenland; greenland sea;} \\
   & \footnotesize{east greenland} & \\ 
\noalign{\smallskip}
36 & \footnotesize{{\em svalbard}; arctic; climate change; spitsbergen; alpine populations; mites} & \\ 
\noalign{\smallskip}
37 & \footnotesize{{\em pollution}; chemistry; instruments and techniques} & \\ 
\noalign{\smallskip}
27 & \footnotesize{{\em amphipods}; weddell sea; optical properties} & \\ 
\noalign{\smallskip}
38 & \footnotesize{{\em rates}; propagation; mountains} & \\ 
\noalign{\smallskip}\hline
\end{tabular}
\end{table}
\begin{table}
\centering
\caption{UNIS' research papers 1994-2014: content (keywords) of clusters positioned in the upper left quadrant (above median density, below median centrality) of the centrality-density diagram (see Figure 3). \newline
          Italic terms: see Table 7. }
%
\begin{tabular}{lll}
\noalign{\smallskip}
\hline\noalign{\smallskip}
no. & keywords & \\
\noalign{\smallskip} 
\noalign{\smallskip}\hline\noalign{\smallskip}
1 & \footnotesize{{\em morphology}; taxonomy; glacial survival; rapd phylogeography; rapds;} & \\
  & \footnotesize{markers; origin; amphipoda; revision; fauna} & \\ 
\noalign{\smallskip}
6 & \footnotesize{{\em oxidative stress}; oxyradical scavenging capacity; mytilus galloprovincialis;} & \\
  & \footnotesize{lipid peroxidation; tosc; mid atlantic ridge; hydrothermal vents; sympagic;} & \\ 
  & \footnotesize{hydrogen peroxide; ozone depletion} & \\ 
\noalign{\smallskip}
3 & \footnotesize{{\em interannual variability}; western barents sea; ice production; shelf water;} & \\ 
  & \footnotesize{storfjorden; overflow; dense water; bottom water; atlantic; halocline} & \\ 
\noalign{\smallskip}
2 & \footnotesize{{\em evolution}; sweden; greenland ice sheet; water flow; storglaciaren;} & \\
  & \footnotesize{temperate glaciers; switzerland; tidewater glaciers; velocity; drainage} & \\ 
\noalign{\smallskip}
19 & \footnotesize{{\em thermal regime}; soil; mountain permafrost; tailings; regions;} & \\
  & \footnotesize{glacier surges; exchange; system; mechanism; herbivory} & \\ 
\noalign{\smallskip}
17 & \footnotesize{{\em permafrost}; diffusion; rock glaciers; active layer; holocene;} & \\
  & \footnotesize{temperatures; iceland; flow; thermosphere; layer} & \\ 
\noalign{\smallskip}
14 & \footnotesize{{\em glacier}; nitrogen; carbon; high arctic glacier; microbial diversity;} & \\
  & \footnotesize{ice core; push moraine; bacterial communities; complex; ice sheet} & \\ 
\noalign{\smallskip}
8 & \footnotesize{{\em facies}; delta; petroleum hydrocarbons; sediment; deposits;} & \\
  & \footnotesize{systems; toxicity; models; tertiary; boundary} & \\ 
\noalign{\smallskip}\hline
\end{tabular}
\end{table}
\begin{table}
\centering
\caption{UNIS' research papers 1994-2014: content (keywords) of clusters positioned in the lower right quadrant (below median density, above median centrality) of the centrality-density diagram (see Figure 3).  \newline
          Italic terms: see Table 7. }
%
\begin{tabular}{lll}
\noalign{\smallskip}
\hline\noalign{\smallskip}
no. & keywords & \\
\noalign{\smallskip} 
\noalign{\smallskip}\hline\noalign{\smallskip}
16 & \footnotesize{{\em sea ice}; gammarus wilkitzkii; crustacea; apherusa glacialis; strategies;} & \\
  & \footnotesize{deep sea; macrobenthos; invertebrates; communities; contaminants} & \\ 
\noalign{\smallskip}
11 & \footnotesize{{\em vegetation}; tundra; island; anser brachyrhynchus; productivity;} & \\
  & \footnotesize{ecosystems; dryas octopetala; moss; population; canada} & \\ 
\noalign{\smallskip}
30 & \footnotesize{{\em magnetosphere}; reconnection; magnetopause; signatures; boundary layer;} & \\
  & \footnotesize{cusp; convection; latitude boundary layer; low altitude; magnetosheath} & \\ 
\noalign{\smallskip}
29 & \footnotesize{{\em life history}; lipids; disturbance; alaska; ungulate; foraging ecology;} & \\
  & \footnotesize{ecology; interspecific competition; community structure; gamasida} & \\ 
\noalign{\smallskip}
25 & \footnotesize{{\em north atlantic}; norwegian sea; diversity; ocean; sea; last glacial period;} & \\
  & \footnotesize{abundance; surface; water; sea ice cover} & \\ 
\noalign{\smallskip}
23 & \footnotesize{{\em red deer}; white tailed deer; north atlantic oscillation; reproductive effort;} & \\
  & \footnotesize{rangifer tarandus; senescence; cassiope tetragona; reproductive success;} & \\ 
  & \footnotesize{body size; mortality} \\
\noalign{\smallskip}
33 & \footnotesize{{\em stable isotopes}; populations; patterns; herbivores; benthos; community;} & \\
  & \footnotesize{herbivorous copepods; fjord; zooplankton community; identification} & \\ 
\noalign{\smallskip}
34 & \footnotesize{{\em plasma}; magnetopause reconnection; moose alces alces; outflow;} & \\
  & \footnotesize{radar observations; acceleration; region; flows; altitude; simulation} & \\ 
\noalign{\smallskip}\hline
\end{tabular}
\end{table}
\begin{table}
\centering
\caption{UNIS' research papers 1994-2014: content (keywords) of clusters positioned in the upper right quadrant (above median density, above median centrality) of the centrality-density diagram (see Figure 3).  \newline
          Italic terms: see Table 7. }
%
\begin{tabular}{lll}
\noalign{\smallskip}
\hline\noalign{\smallskip}
no. & keywords & \\
\noalign{\smallskip} 
\noalign{\smallskip}\hline\noalign{\smallskip}
5 & \footnotesize{{\em plant phenology}; svalbard reindeer; rangifer tarandus platyrhynchus;} & \\
  & \footnotesize{body weight; density dependence; large herbivores; abomasal nematodes;} & \\
  & \footnotesize{ostertagia gruehneri; marshallagia marshalli; dna evidence} & \\ 
\noalign{\smallskip}  
7 & \footnotesize{{\em heinrich events}; iceberg discharges; north atlantic ocean; western norway;} & \\
  & \footnotesize{last deglaciation; northern north sea; quaternary; glaciation history;} \\
  & \footnotesize{internal structure; fluid flow} & \\
\noalign{\smallskip}  
9 & \footnotesize{{\em interplanetary magnetic field}; flux transfer events; solar wind;} & \\
  & \footnotesize{dayside magnetopause; cutlass finland radar; ionospheric convection;} & \\
  & \footnotesize{dayside auroral activity; moving auroral forms; high latitude magnetopause;} \\
  & \footnotesize{polar cusp} & \\ 
\noalign{\smallskip}  
15 & \footnotesize{{\em ionosphere}; particle precipitation; magnetospheric physics;} & \\
  & \footnotesize{ground based observations; polar ionosphere; incoherent scatter radar;} \\
  & \footnotesize{cusp ion precipitation; eiscat observations;} & \\
  & \footnotesize{auroral phenomena; electric field} & \\ 
\noalign{\smallskip}
20 & \footnotesize{{\em f region}; high latitude convection; polar cap boundary;} & \\
  & \footnotesize{high latitude ionosphere; polar cap; gradient drift instability;} \\
  & \footnotesize{patches; cusp aurora; field; eiscat svalbard radar} & \\ 
\noalign{\smallskip}  
10 & \footnotesize{{\em greenland ice}; nordic seas; rapid changes; thermohaline circulation;} & \\
  & \footnotesize{ocean circulation; late quaternary; northeastern atlantic; deep water;} \\
  & \footnotesize{climate records; foraminifera} & \\ 
\noalign{\smallskip}
4 & \footnotesize{{\em finmarchicus}; calanus glacialis; reproduction; fatty acids; ice algae; life cycle;} & \\
  & \footnotesize{food quality; trophic relationships; fatty acid composition; energy} & \\ 
\noalign{\smallskip}
13 & \footnotesize{{\em barents sea}; zooplankton; marginal ice zone; ecosystem;} & \\
  & \footnotesize{northeast water polynya; taylor valley; beaufort sea; organic matter;} \\ 
  & \footnotesize{biomass; seasonality} & \\ 
\noalign{\smallskip}
12 & \footnotesize{{\em phenology}; stratigraphy; norway; responses; growing season; suez rift;} & \\
  & \footnotesize{plants; central spitsbergen; deformation; sedimentation} & \\ 
\noalign{\smallskip}
24 & \footnotesize{{\em arctic ocean}; circulation; fram strait; kongsfjorden; waters; atlantic water;} & \\
  & \footnotesize{copepods; west spitsbergen current; phytoplankton; mass} & \\ 
\noalign{\smallskip}
18 & \footnotesize{{\em population dynamics}; cervus elaphus; density; age; body mass; ungulates;} & \\
  & \footnotesize{roe deer; sex ratio; tooth wear; bighorn sheep} & \\ 
\noalign{\smallskip}
\hline\noalign{\smallskip}
\end{tabular}
\end{table}

\end{document}